# Modeling dynamic surface tension on surfactant-enhanced polydimethylsiloxane (PDMS)

Daniel Joseph O'Brien and Makarand Paranjape

**Abstract**

*Hypothesis*
Surfactants are often added to aqueous solutions to induce spreading on otherwise unwettable hydrophobic surfaces. Alternatively, they can be introduced directly into solid hydrophobic materials—such as the soft elastomer, polydimethylsiloxane—to induce autonomous spreading without requiring additional surface or liquid modifications. Given the similarity between mechanisms of these two approaches, models that describe wetting by aqueous surfactant solutions should also characterize wetting on surfactant-solid systems.

*Experiments*
Multiple surfactants of varying size and chemical composition were added to pre-polymerized PDMS samples. After cross-linking, water droplets were placed on the surfaces at set time points, and their contact angles were recorded to track the temporal evolution of the interfacial tension. Multiple nonlinear models were fitted to this data, their parameters analyzed, and each goodness of fit compared.

*Findings*
An empirical model of dynamic surface tension was found to describe the wetting process better than the single established model found in the literature. The proposed model adapted better to the longer timescales induced by slow molecular diffusivity in PDMS. Siloxane ethoxylate surfactants induced faster and more complete wetting of PDMS by water than oxyoctylphenol ethoxylates did. The generalizability of this model for characterizing surfactants of a wide range of physiochemical properties was demonstrated.

## 1. Introduction

Dynamic surface tension (DST) is the process by which surfactant molecules in solution migrate to an interface in order to decrease the interfacial tension and induce wetting. The kinetics of the process are determined by the diffusion and adsorption rates of the surfactant. These kinetics have been studied in detail for systems where an aqueous surfactant solution spreads on an originally hydrophobic surface [1,2]. In such systems, wetting occurs quickly because of the rapid diffusion of solute in water. DST processes have long been utilized in industry, such as for enhancing the spreading of agricultural pesticides [3] or for the production of thin films for photographic applications [4].

Alternatively, if one desires to exploit the advantageous properties of such an interface without modifying the wetting liquid, surfactants can instead be added to the solid material rather than to the liquid that wets it. This technique is most widely employed by embedding surfactants in silicones. For example, surfactants have been used to improve the wetting of silicone dental impression materials over tooth structures [5], to generate fouling-release coatings for protecting seawater-exposed materials [6,7], to increase the lubricity of films in mechanical systems [8], and in PDMS-based microfluidics for inducing autonomous capillary flow [9] and combating non-specific protein adsorption [10].

Despite the many applications of DST, surfactant diffusion and adsorption kinetics in silicones are not well studied compared to the kinetics of aqueous solutions of surfactant. Recently, Starov's theory of surfactant adsorption was adopted to model the wetting of porous, surfactant-laden PDMS [11,12]. However, this investigation was limited to a single trisiloxane ethoxylate surfactant (Silwet L-77) and the model is not applicable for describing dilute mixtures. In this paper, we instead adapt Hua and Rosen's DST model [13], which was derived empirically and improves upon Starov's model in its compatibility with the longer timescales of surfactant diffusion in silicones. Here, our model demonstrates its generalizability to surfactants of multiple chemistries, diffusion coefficients, molecular weights, and hydrophilic-lipophilic balances. As the choice of the most effective surfactant is critical to efficient wetting [14], this model could assist researchers across a wide range of fields to properly select surface-active molecules for their applications.

**2. Theory**

2.1 The stages of dynamic surface tension
Hua and Rosen first detailed the evolution of surface tension data for aqueous surfactant solutions on hydrophobic surfaces [13]. Their analysis yielded a graphical plot of surface tension as a function of time evolution and suggested that the process occurs in four distinct stages: an initial, stable induction phase, followed by a fast fall, then a slowly stabilizing mesoequilibrium phase, finally culminating in a steady-state equilibrium phase. In this paper, the cosine of the contact angle is plotted with time, as in our previous work [15] and shown in Fig. 1, rather than Hua and Rosen's surface tension parameter, with the two being inversely related to one another. Therefore, Hua and Rosen's "fast fall" phase is renamed "rapid rise."

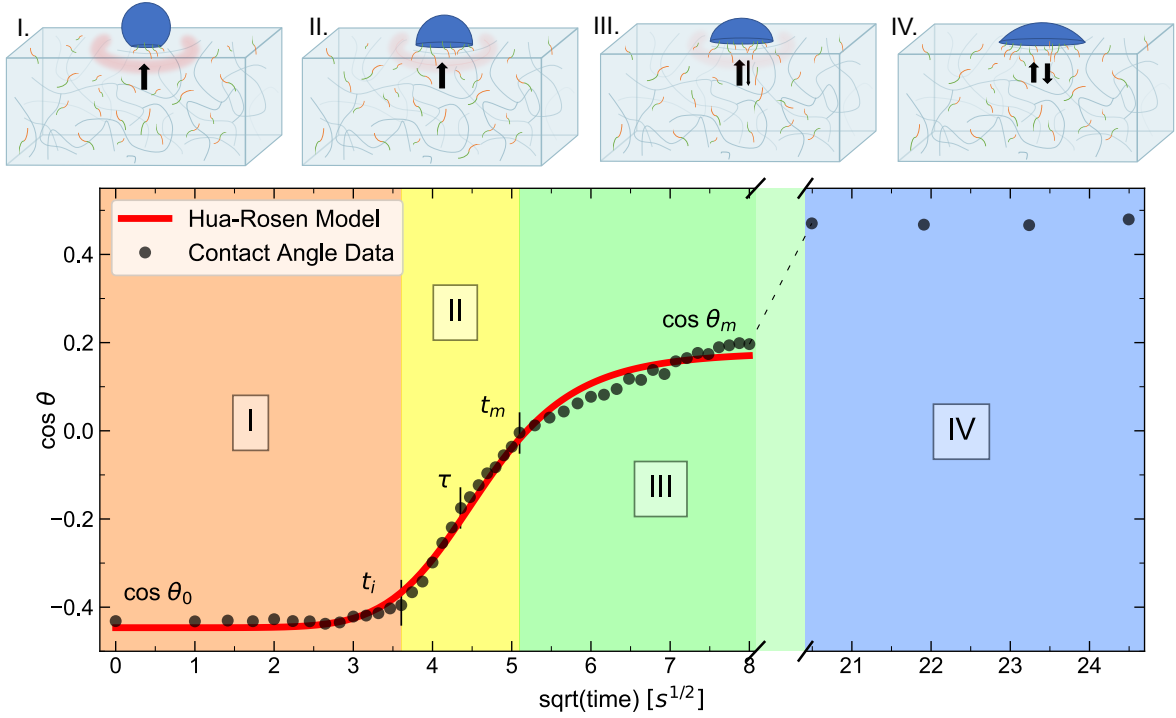

**Fig. 1.** The four stages of DST demonstrated with surfactant-enhanced PDMS: I. induction; II. rapid rise (called "fast fall" by Hua and Rosen); III. mesoequilibrium; and IV. equilibrium. Diffusion-limited adsorption of surfactant at the PDMS-water interface is driven by a concentration gradient under the wetted area (red halo). Data represents the average of three PDMS samples with the surfactant Silwet L-77 at 0.6 wt%.

During the induction stage, surfactant from the PDMS subsurface (immediately below the wetted area) adsorbs to the interface, inducing a concentration gradient within the bulk. Although surfactant is adsorbing at the surface, little visible change occurs until approximately 2/3 of the surface is filled [16]. Lin attributed this slow relaxation phase to strong intermolecular attraction between adsorbed species [17]. These interactions lead to a gas-liquid expanded phase transition of the adsorbing surfactant, at the conclusion of which surface tension changes dramatically [18]. This change marks the beginning of the rapid rise stage, during which the contact angle changes quickly as surface tension decreases. The slope of $\cos\theta$ vs. $t^{1/2}$ during this stage can be used to calculate the surfactants' diffusivity, as will be explained in Sec. 2.2. At $t_m$, the system enters the mesoequilibrium stage, when diffusion of adsorbed surfactant molecules back into bulk PDMS becomes non-negligible and the change in surface tension slows. Finally, at long $t$, the system reaches the equilibrium phase when the surface has been filled and solute diffusive processes between the interface and bulk phases equilibrate.

## 2.2 Diffusivity

For the case being considered, Fick's second law in one dimension,

$$\frac{\partial c(x,t)}{dt} = D\frac{\partial^2 c(x,t)}{\partial x^2}, \tag{1}$$

governs the process of diffusion. Here, $D$ represents the diffusion coefficient, or diffusivity, of the diffusant, in dimensions of area per unit time. Using appropriate boundary conditions, Ward and Tordai [19] solved this equation to find the time-dependent surface excess of surfactant adsorbed at a newly formed interface as

$$\Gamma(t) = 2c_0 \left(\frac{Dt}{\pi}\right)^{1/2} - 2\left(\frac{D}{\pi}\right)^{1/2} \int_0^{\sqrt{t}} c_s d(\sqrt{t-\tau}), \tag{2}$$

where $c_0$ is the bulk surfactant concentration, $c_s$ is the subsurface concentration, and $\tau$ is a dummy integration variable. For short time scales, the second term on the right of the equality, representing backwards diffusion from the interface into the bulk, can be neglected. It has been shown that if one assumes diffusion-limited adsorption, at short times, DST can be characterized by the evolving contact angle, θ, such that

$$\cos\theta = \frac{2c_0 RT}{\gamma_{LV}} \sqrt{Dt/\pi} + \frac{\gamma_{SL_0} - \gamma_{SV}}{\gamma_{LV}}, \tag{3}$$

where $R$ is the gas constant, $T$ the temperature, and $\gamma_{LV}$, $\gamma_{SV}$, and $\gamma_{SL_0}$ are the liquid-vapor, solid-vapor, and initial solid-liquid surface tensions [15,20]. Using this relationship, the surfactant diffusivity, $D$, can be extracted using the slope of the linear change in $\cos\theta$ vs. $\sqrt{t}$ during the rapid rise stage,

$$D = \pi \left[\left(\frac{\partial \cos\theta}{\partial t^{1/2}}\right) \frac{\gamma_{LV}}{2c_0 RT}\right]^2. \tag{4}$$

2.2 Starov DST Model

Per Starov, the time-dependent wetting of a surface by an aqueous surfactant solution can be described as

$$\cos\theta(t) = \cos\theta_0 - (\cos\theta_0 - \cos\theta_\infty)(1 - e^{-\frac{t}{\tau}}), \tag{5}$$

where $\theta_0$ and $\theta_\infty$ are the initial and final contact angles of the liquid and $\tau$ is a parameter called the surfactant transfer time [12]. It has been shown that this process is rate-limited by the transfer of aqueous surfactant molecules onto the solid-vapor interface in front of the spreading drop [21].

2.3 Hua-Rosen DST Model

Alternatively, interfacial tension over the first three stages of DST can be modeled as

$$\gamma_t - \gamma_m = \frac{\gamma_0 - \gamma_m}{1 + \left(\frac{t}{\tau}\right)^n}, \tag{6}$$

where $\gamma_0$ is the solid-liquid surface tension at $t = 0$, $\gamma_t$ is at any time t, and $\gamma_m$ is at mesoequilibrium. The parameter $\tau$ has dimensions of time while $n$ is a dimensionless quantity [13]. When $t = \tau$, the surface pressure of surfactant, $\Pi = \gamma_0 - \gamma_\tau$, is equal to half its value at mesoequilibrium; that is

$$\Pi_\tau = \gamma_0 - \gamma_\tau = \frac{1}{2}\Pi_m = \frac{1}{2}(\gamma_0 - \gamma_m). \tag{7}$$

This situation occurs during the rapid rise stage, where the slope of $\cos\theta$ vs. $t^{1/2}$ is at its maximum and can be used to calculate surfactant diffusivity. Using Young's equation, which relates the surface tensions in a planar, three-phase system as $\gamma_{SL} - \gamma_{SV} + \gamma_{LV}\cos\theta = 0$, contact angle can be substituted for surface tension in Eq. (6) to yield

$$\cos\theta(t) = \cos\theta_m - \frac{\cos\theta_m - \cos\theta_0}{1 + \left(\frac{t}{\tau}\right)^n}, \tag{8}$$

which is of similar form to Starov's model, but with a power law-dependence in place of an exponential decay. This model was developed empirically, and some physical parameters can be derived from it, as will be explored in this paper for PDMS-surfactant systems.

## 3. Materials and methods

3.1 Materials
The PDMS used is a Sylgard 184 silicone elastomer formulation and was used as received. All surfactants were also used as received and are presented alongside their molecular weights (MW) and hydrophilic-lipophilic balances (HLB) in Table 1. Their chemical structures are shown in Fig. 2.

**Table 1:** Surfactants categorized by chemical structure, molecular weight (MW), and hydrophilic-lipophilic balance (HLB).

| Chemical Structure | Name | MW (g/mol) | HLB | Supplier |
|---|---|---|---|---|
| Poly(dimethylsiloxane-ethylene oxide) block copolymer | PDMS-b-PEO | 600 | 15 | Polysciences |
| | DBE-311 | 1000 | 6.5 | Gelest |
| | DBE-411 | 450 | 9.5 | Gelest |
| | DBE-712 | 600 | 13 | Gelest |
| | DBE-814 | 1000 | 16 | Gelest |
| Octylphenol ethoxylate | Triton X-45 | 404 | 9.8 | Fisher |
| | Triton X-114 | 537 | 12.4 | Fisher |
| | Triton X-100 | 625 | 13.5 | Fisher |
| | Triton X-102 | 757 | 14.4 | Fisher |
| Trisiloxane ethoxylate | Silwet L-77 | 339 | 12 | Momentive |

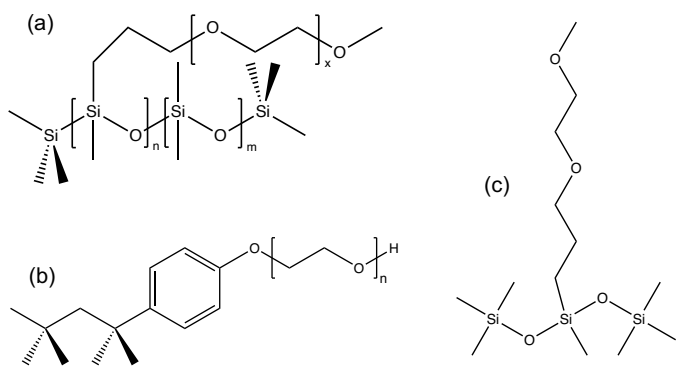

**Fig. 2.** Diagram of the chemical structures of surfactants used. (a) PDMS-PEO block co-polymers, (b) Triton octylphenol ethoxylates, and (c) Silwet L-77 trisiloxane ethoxylate.

3.2 Sample preparation
Sylgard 184 was mixed in a 10:1 base:curing agent ratio. Between 0.2–3.0 wt% of the desired surfactant was then added, and the three-part blend was stirred for ten minutes. All samples were then degassed to remove bubbles, then cured for 4.5 hours at 65 °C.

3.2 Contact angle measurements and analysis
Images of 5 µL DI $H_2O$ droplets on each surfactant-PDMS sample were recorded with a Basler acA4096-30um camera and a 10X close focus zoom lens (Edmund). Contact angles were calculated with DropSnake polynomial fitting on Fiji [22,23] and the data was fitted with both the Starov and Hua-Rosen DST models using OriginLab Pro. Relevant parameters were extracted and analyzed, and diffusivities were calculated per the method in Sec. 2.1. The two models were quantitatively compared utilizing the Akaike information criterion (AIC) and Bayesian information criterion (BIC) in OriginLab Pro. These statistical tests quantitatively compare the fit of nonlinear models to data, weighing goodness of fit against the number of parameters in the models.

## 4. Results and discussion

4.1 Model comparison
Surfactants are able to diffuse rapidly in aqueous solution. For example, Hua and Rosen's measurements of $\tau$, the time for surfactant at an interface to reach ½ its final surface pressure, ranged between $10^{-1}$ and $10^1$ s. In contrast, surfactants in solid PDMS networks diffuse slowly. As a result, the time required for a large proportion of the interface to be filled with surfactant is long, and therefore the induction period lasts longer in these systems. This delay can be shortened by increasing the concentration of surfactant or by utilizing amphiphiles with faster diffusivities.

Due to the longer timescales involved, Starov's model has the limitation that it only fits a narrow range of surfactant-PDMS systems prepared with certain parameters. Due to its exponential dependence, this model is unable to account for the long induction times found in dilute PDMS-surfactant systems. Previously, the model has only been utilized to model Silwet L-77 surfactant's effect on PDMS wetting, and at lower concentrations (0.5 to 1.0 wt%), some data deviated significantly from the model [11]. Alternatively, the power law-dependence of the Hua-Rosen model allows more generalizable data analysis. Figure 3 shows a comparison of the two models' fits to experimental data using three popular surfactants, Triton X-100, PDMS-b-PEO, and Silwet L-77. These surfactants vary in chemical structure; Triton is an octylphenol ethoxylate, PDMS-b-PEO is a comb-like siloxane ethoxylate, and Silwet L-77 is a trisiloxane ethoxylate. Whereas the Hua-Rosen model was able to fit all data sets, Starov's model failed to converge when fitting data of Triton X-100 samples with c < 1.5 wt%. The AIC and BIC statistical tests determined that for all data sets, the Hua-Rosen equation was the preferred model (statistics available in S1). As such, for the remainder of this manuscript, only the Hua-Rosen model will be explored.

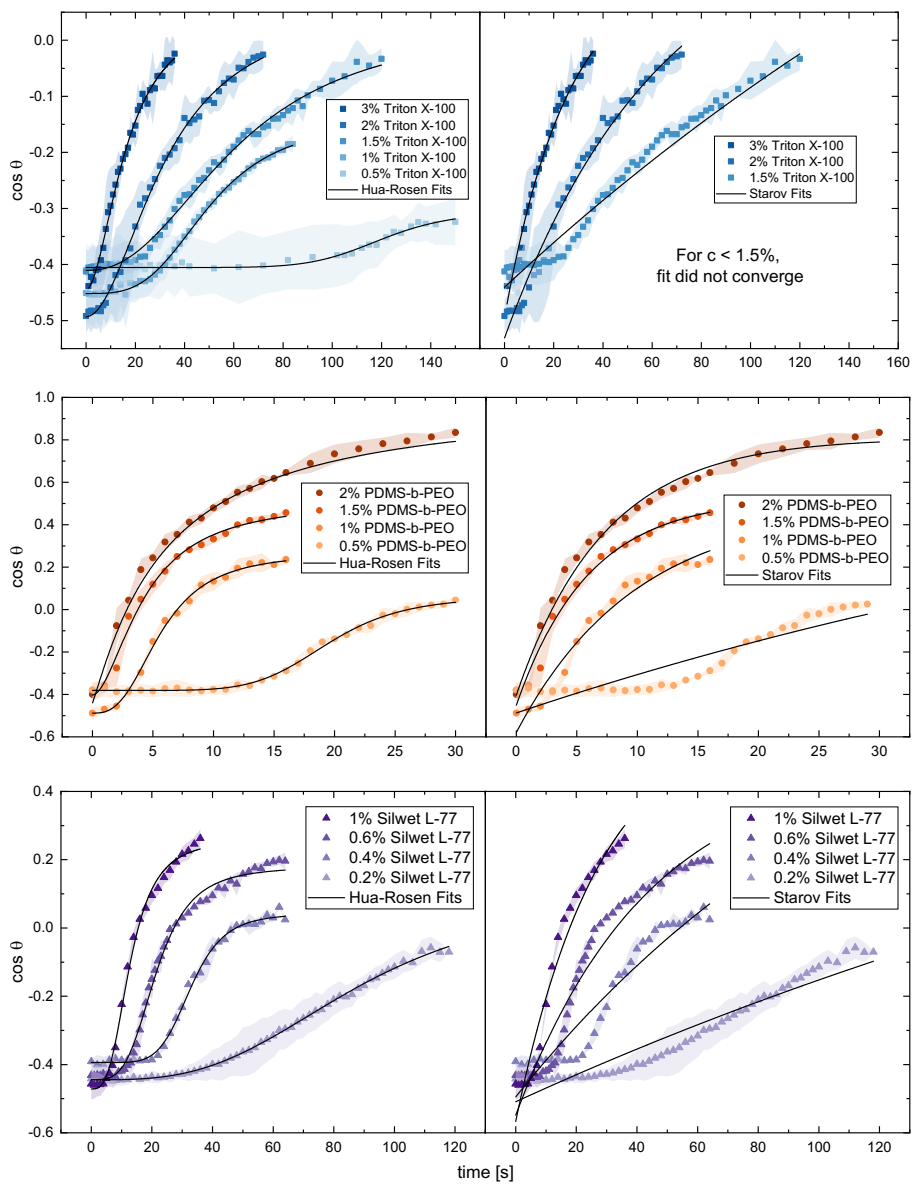

**Fig. 3.** Comparison of Hua-Rosen (*left*) and Starov (*right*) fits for multiple concentrations (wt%) of three surfactants in PDMS: Triton X-100 (*top*), PDMS-b-PEO (*middle*), and Silwet L-77 (*bottom*). The data sets are the same for the two models under test.

## 4.2 Concentration effects

The effects of concentration, chemical structure, and size on Hua-Rosen model parameters are demonstrated in Figure 4. As shown in Figure 4(a), $\log \tau$, which defines the timescale for effective wetting of a sample (Eq. 8), varies linearly with increase in $\log c$ (over a wide range from ~5 s to over 100 s). Consequently, it is important to understand the advantages and disadvantages of utilizing high concentrations of surfactants in PDMS systems. For example, high concentrations of Triton X-100 both increase PDMS opacity [24] and have a plasticizing effect, softening PDMS and decreasing its tensile modulus [25]. High concentrations of PDMS-PEO can impede the ability of PDMS to bond to glass or silicon, as is regularly required for microfluidic systems [9]. Concentrations above 1 wt% of Silwet can make PDMS degassing difficult [26].

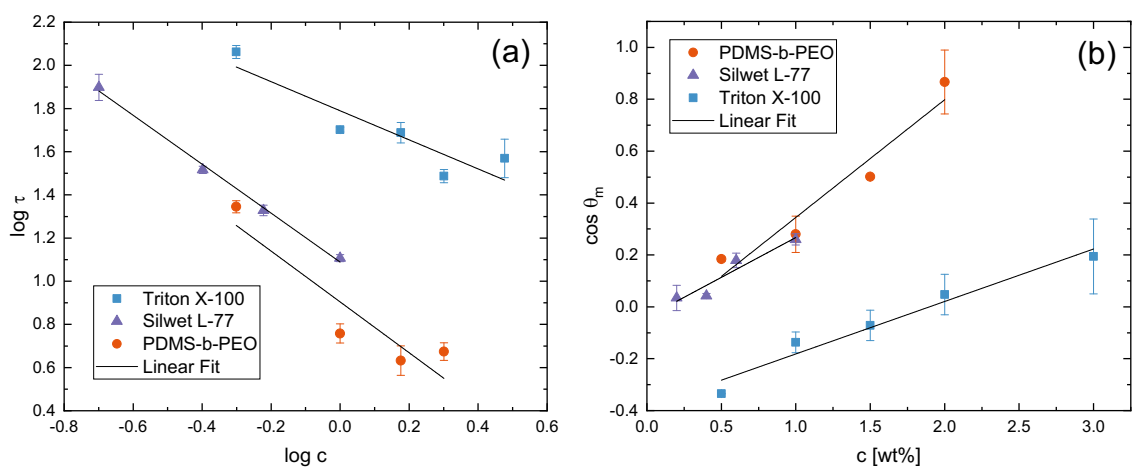

**Fig. 4.** Characterizing wetting through fit parameters. (a) Plot of the logarithm of $\tau$ (Hua-Rosen parameter) vs. logarithm of concentration for three commonly used surfactants. (b) Plot of $\cos \theta_m$ vs. concentration of the same surfactants.

It is also evident in Figure 4(a) that chemical composition is crucial to the surfactant's efficacy. Silwet and PDMS-b-PEO surfactants, both having flexible siloxane hydrophobic moieties, follow similar trends for wetting, with their trendlines nearly overlapping. Triton X-100, on the other hand, features a large phenyl group, which induces steric hindrance to diffusion. As a result, the timescale for wetting is significantly longer.

Figure 4(b) shows that cosine of $\theta_m$, the mesoequilibrium contact angle, increases linearly with concentration. Similar to the trend in timescales, siloxane-based amphiphiles are also shown to provide more complete wetting than Triton X-100, even with concentrations as low as 0.2 wt%.

No significant trends in $n$, the last Hua-Rosen parameter, were noted with concentration variation.

## 4.3 Surfactant size effects

It has been shown that siloxane-based amphiphilic molecules can diffuse at least ten times more quickly through PDMS than octylphenol-based amphiphiles [15]. In order to study the effect of surfactant size, multiple surfactants of similar chemical compositions were investigated. As seen in Table 1, four Triton surfactants with molecular weights ranging from 404–757 g/mol and five comb-like PDMS-PEO surfactants with weights ranging from 450–1000 g/mol were compared. Whereas the HLB of Triton samples correlated with size, the PDMS-PEO samples were more varied (Fig. 5(a)).

From Equation 1, the diffusion timescale should scale as $\tau = L^2/D$, where $L$ is the length scale of the system and $D$ is the diffusant's coefficient. The time constant $\tau$ of each surfactant, extracted from the Hua-Rosen model, was plotted against its diffusivity, calculated with the slope of $\cos\theta$ vs. $t^{1/2}$ in the rapid rise phase of wetting. As shown in Figure 5(b), the two were inversely correlated, supporting the assumption of diffusion-limited adsorption and therefore validating the diffusivity calculation method.

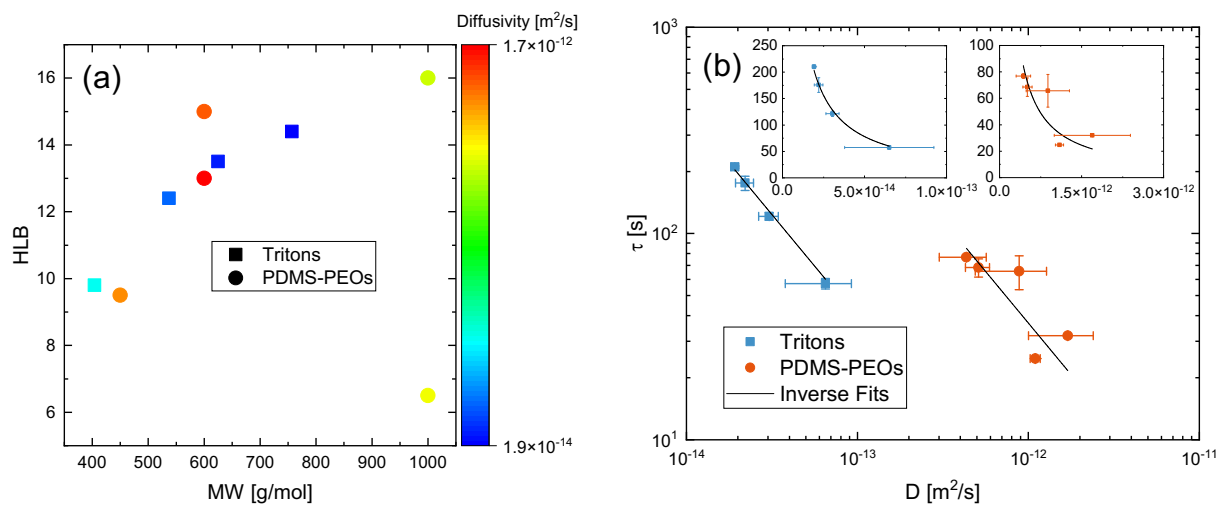

**Fig. 5.** Validating diffusion-limited adsorption and diffusivity calculation. (a) Gradient-colored scatter plot of surfactant HLB vs. MW with diffusion coefficients represented by color, and (b) log-log plot and linear inset plots show the inverse relation between $\tau$ and $D$, the diffusivity of the surfactant.

## 5. Conclusions
We have adapted a model from the literature to characterize dynamic surface tension induced by the introduction of surfactants into solid silicone PDMS, in contrast to its addition into aqueous solution. Fitting the model to contact angle data and examining the parameters outputs, we were able to demonstrate how surfactant physiochemical properties affected the timescale and completeness of wetting that the amphiphiles induce. The results supported our previously introduced method for calculating the diffusivity of surfactants in PDMS [15]. To date, only one model had been proposed for describing surfactant-PDMS DST. However, because of the long timescales for adsorption in this system, that model would only fit a narrow window of system preparations [11]. In contrast, the proposed model is extremely generalizable, describing well the adsorption of multiple surfactants of varying size, chemical composition, and concentration. This model can be used to assist researchers in many fields in the proper selection of surfactants for their application.

**Acknowledgements**

This work was supported by the National Science Foundation (# 1938995); and the Achievement Rewards for College Scientists (ARCS) – Metro Washington Chapter. The authors wish to thank P. Olmsted for insightful discussions on diffusion, as well as P. Bhatia, J. Farina, S. Fisher, and D. Mills for their review of the manuscript.